\RequirePackage[2024-06-01]{latexrelease}
\documentclass[aps,prb,preprint,superscriptaddress]{revtex4-2}
\usepackage{dcolumn}   
\usepackage{bm}        
\usepackage{amsmath}
\usepackage{amssymb}
\usepackage[utf8]{inputenc}
\usepackage{graphicx,siunitx}
\usepackage{amsmath}
\DeclareUnicodeCharacter{2212}{-}
\usepackage{booktabs}
\usepackage{subcaption}
\usepackage[svgnames]{xcolor}
\usepackage[export]{adjustbox}
\usepackage[percent]{overpic}
\usepackage{array}
\usepackage{float}
\usepackage[style=base]{caption}

\begin{document}

	\title{Pressure induced structural transitions, coordination crossover, and optoelectronic response in divalent metavanadate CaV$_2$O$_6$}
	
    \author{Bhagyashri Giri}
    \affiliation{Department of Physical Sciences, Indian Institute of Science Education and Research Kolkata, Mohanpur Campus, Mohanpur 741246, Nadia, West Bengal, India.}
	\affiliation{National Centre for High-Pressure Studies, Department of Physical Sciences, Indian Institute of Science Education and Research Kolkata, Mohanpur Campus, Mohanpur 741246, Nadia, West Bengal, India.}
	\author{Asish Kumar Mishra}
	\affiliation{Department of Physical Sciences, Indian Institute of Science Education and Research Kolkata, Mohanpur Campus, Mohanpur 741246, Nadia, West Bengal, India.}
	\affiliation{National Centre for High-Pressure Studies, Department of Physical Sciences, Indian Institute of Science Education and Research Kolkata, Mohanpur Campus, Mohanpur 741246, Nadia, West Bengal, India.}
	\author{Bidisha Mukherjee}
	\affiliation{IMPMC, Sorbonne Université, CNRS,	MNHN, 4, place Jussieu, 75005 Paris, France.}
	\author{Debabrata Samanta}
	\affiliation{Beijing Academy of Quantum Information Sciences, Beijing, P.R. China.}
	\author{Nico Giordano}
	\affiliation{Photon Science, Deutsches Elektronen Synchrotron, 22607 Hamburg, Germany.}
    \author{Goutam Dev Mukherjee}
	\email [Corresponding author: ]{goutamdev@iiserkol.ac.in}
	\affiliation{Department of Physical Sciences, Indian Institute of Science Education and Research Kolkata, Mohanpur Campus, Mohanpur 741246, Nadia, West Bengal, India.}
	\affiliation{National Centre for High-Pressure Studies, Department of Physical Sciences, Indian Institute of Science Education and Research Kolkata, Mohanpur Campus, Mohanpur 741246, Nadia, West Bengal, India.}
	\date{\today}
	
\begin{abstract}
High-pressure synchrotron X-ray diffraction, Raman spectroscopy, photoluminescence (PL), and UV-Vis absorption measurements were carried out on the divalent metavanadate CaV$_2$O$_6$ up to 31 GPa at room temperature. Ambient CaV$_2$O$_6$ crystallizes in a centrosymmetric pseudo-brannerite monoclinic structure ($C2/m$), where vanadium atoms reside in fivefold-coordinated VO$_5$ trigonal bipyramids. Under hydrostatic compression, the system undergoes a structural phase transition starting at $\sim$1 GPa to a lower-symmetry, non-centrosymmetric monoclinic phase ($C2$), which becomes the single stable phase above 5.2 GPa. This transition involves an increase in vanadium coordination from VO$_5$ bipyramids to edge-sharing distorted VO$_6$ octahedra. High-pressure Raman scattering corroborates the structural phase boundaries, exhibiting mode softening ($\gamma_3 = -0.42$, $\gamma_9 = -0.46$) within the mixed-phase regime (1.0 - 4.4 GPa) and sharp anomalies in mode frequencies and linewidths across 5.2 GPa. The optical measurements show complete PL quenching near 1 GPa due to non-radiative recombination pathways introduced by local polyhedral distortions, alongside a non-monotonic evolution of the indirect optical band gap. Above 20 GPa, an anomaly in polyhedral distortion and internal strain drives an isosymmetric structural modification within the $C2$ phase, marked by an alteration in bulk compressibility and stabilization of the optical gap. Comparative high-pressure band-gap measurements on the sister compound MgV$_2$O$_6$ validate the explicit coupling between cation coordination change and optoelectronic response in divalent metavanadates. 
\end{abstract} 

\maketitle
	
\section{INTRODUCTION}
Since the discovery of the metal-insulator transition (MIT) in V$_2$O$_3$ \cite{morin1959oxides,austin1962effect}, vanadium oxide-based materials have attracted significant interest due to their diverse structural, electronic, magnetic, and optical properties  \cite{li2017pressure,weckhuysen2003chemistry,arora2012high}. Depending upon the valence states of V, vanadium oxides stabilize in different crystal structures: VO (rocksalt structure), V$_2$O$_3$ (corundum structure), VO$_2$ (rutile structure), V$_2$O$_5$ (layered orthorhombic structure) \cite{blum2007surface}. For decades, researchers have shown keen interest in exploring the vanadium oxide based materials under extreme conditions of pressure and temperature in an effort to understand their diverse physical properties. Temperature-dependent MIT is observed in many vanadium oxide systems such as VO$_2$, V$_2$O$_3$, and V$_4$O$_7$ \cite{biermann2005dynamical,dernier1970crystal,botana2011electronic}. Arcangeletti et al. \cite{arcangeletti2007evidence} reported a pressure-induced MIT in monoclinic VO$_2$ above 10 GPa, evidenced by band gap filling in optical spectra while the monoclinic structure remained unchanged. On the other hand, with increasing pressure up to 32.5 GPa, the metallic phase of V$_2$O$_3$ turns into an insulating phase with increasing resistivity \cite{ding2014novel}. At around 4 GPa, two-dimensional layered orthorhombic $\alpha$-V$_2$O$_5$ powder consisting of VO$_5$ square pyramids transforms into three-dimensional monoclinic $\beta$-V$_2$O$_5$ with edge sharing VO$_6$ octahedra, accompanied by a decrease in resistivity from 10000 ohm.cm for $\alpha$-V$_2$O$_5$ to 400 ohm.cm for $\beta$-V$_2$O$_5$ \cite{gallardo2007computational,filonenko2004structure}. Thus, the wide variety of structural and electronic changes in vanadium oxides has led scientists to delve deeper into studying various vanadium oxide-based compounds under high-pressure. Alkali metavanadates MVO$_3$ (M= Li, K, Rb, Cs) undergo structural phase transitions at high-pressure \cite{adams1988vibrational,shen1995high,kourouklis1991high}. For example, LiVO$_3$ exhibits a first-order phase transition around 4 GPa, accompanied by a complete change in Raman spectra, followed by two additional second-order phase transitions at higher pressure \cite{adams1988vibrational}. $\alpha$-NaVO$_3$ undergoes pressure-induced amorphization at around 6.1 GPa, accompanied by the disappearance of Raman modes and emergence of broad bands \cite{shen1994pressure}. Also, a series of quasi-1D metallic $\beta$-vanadium bronzes $\beta$-A$_{0.33}$V$_2$O$_5$ (A= Li$^+$, Na$^+$, Ag$^+$, Ca$^{2+}$, Sr$^{2+}$, Pb$^{2+}$), all having a mixed valence state V$^{4+}$(3d$^1$) and V$^{5+}$(3d$^0$), show charge ordering (CO) type MIT (except Pb$^{2+}$) with lowering temperature and at ambient pressure. Among them, in all A$^+$ compounds, the CO phase collapses with increasing pressure and drives the system to a superconducting phase \cite{yamauchi2007rich}. $\beta$-Na$_{0.33}$V$_2$O$_5$ is the first known superconductor in the family of $\beta$-vanadium bronzes, exhibiting superconductivity at approximately 6K and under a pressure of 8  GPa \cite{yamauchi2002pressure}. For divalent A$^{2+}$ compounds, superconductivity is not observed, but the CO phase is suppressed at about 3.5 GPa for Ca$^{2+}$ and 1.5 GPa for Sr$^{2+}$, respectively \cite{yamauchi2007multiple,yamauchi2007rich}. Thus, from simple vanadium oxides to alkali vanadates, orthovanadates, and $\beta$-vanadium bronzes, all are promising materials for their wide variety of structural, electronic, and magnetic properties at extreme conditions of pressure and temperature.

Within this perspective, the divalent metavanadate family MV$_2$O$_6$ (M=Ca, Mg, Zn, Ba, Mn, Ni, Cu, etc.) is considered to be a promising candidate for further exploration. The metavanadates have received considerable attention due to their applications in various fields like optoelectronic devices, rechargeable batteries, catalysts, dielectric ceramics, etc. \cite{lakshminarasimhan2022optical,abhiram2025comprehensive,vaz2020low}. These materials are classified into four different structures: Columbite, Brannerite, Pseudo-brannerite, and NiV$_2$O$_6$ type structure \cite{mocala1987polymorphism}. For the ideal brannerite structure (M=Mg, Zn, Cu etc), M cation is in six fold coordination, and V is also in six (5+1) fold coordination with oxygen atoms, forming MO$_6$ and VO$_6$ octahedra. But for the pseudo-brannerite structure, V is in five fold coordination with oxygen atoms, forming a bipyramidal geometry.  MgV$_2$O$_6$, having VO$_6$ octahedra, undergoes a coordination change at about 4 GPa, followed by structural phase transition from $C2/m$ to $C2$ phase at around 20 GPa. The two phases coexist up to the highest pressure studied, along with a two orders of magnitude decrease in resistivity above the structural transition \cite{tang2016exploring}. A sister compound, ZnV$_2$O$_6$, also undergoes phase transition from $C2/m$ to $C2$ at around 16 GPa, along with a change in slope of resistivity around the same pressure \cite{tang2014reversible}. In a comparative high-pressure study up to 15 GPa of Zn-vanadates ZnV$_2$O$_6$, Zn$_2$V$_2$O$_7$, and Zn$_3$V$_2$O$_8$, Díaz-Anichtchenko et al.\cite{diaz2020comparative} have shown that, Zn$_2$V$_2$O$_7$ undergoes three phase transitions at about 0.3 GPa, 3.8 GPa and above 10.8 GPa. SrV$_2$O$_6$ and BaV$_2$O$_6$ consist of VO$_4$ tetrahedral units, undergo pressure-induced structural transition around 3.9 GPa and 4.3 GPa, respectively, followed by pressure-induced amorphization at around 10 GPa \cite{li2015pressure}. Consistent with these findings, high-pressure XRD study by Zhang et al. revealed that BaV$_2$O$_6$ undergoes a structural phase transition from orthorhombic C222 phase to monoclinic C2 phase at around 4 GPa \cite{zhang2025high}.

At ambient condition, CaV$_2$O$_6$ possesses pseudo-brannerite structure consisting of edge sharing CaO$_6$ octahedra and VO$_5$ bipyramidal structure. The CaO$_6$ octahedra and VO$_5$ pyramids are interconnected with each other by corner sharing  \cite{yu2015structure}. It is an indirect semiconductor with an optical band gap of 2.52 eV \cite{yu2015structure}. The valence band and conduction band primarily originate from  O-2p and V-3d orbitals, respectively as a result of crystal field  splitting of V-O polyhedra \cite{jin2022comprehensive}.  CaV$_2$O$_6$ exhibits photoluminescence with a broad emission centered at about 705 nm and a large Stokes shift of 1.15 eV. Thus the unique pseudo-brannerite structure, optical luminescence properties make CaV$_2$O$_6$ different from other metavanadates. Also, a careful look upon other metavanadates shows that the reported structural transition pressure of divalent metavanadate MV$_2$O$_6$ having space group $C2/m$, have also a correlation with the ionic radius of M$^{2+}$ cation; for example, the transition pressure decreases from $\sim$ 20 GPa for MgV$_2$O$_6$ (r=0.72 \AA) to $\sim$ 16.6 GPa for ZnV$_2$O$_6$ (r=0.74 \AA) \cite{tang2016exploring, tang2014reversible}. Therefore, having a larger ionic radius of Ca$^{2+}$ (r=1.0 \AA), it will be worthwhile  to investigate the effect of external pressure on the structural, electronic, and optical properties of CaV$_2$O$_6$.

In the present work, we have investigated the pressure-dependent structural, vibrational, and optical properties of CaV$_2$O$_6$ using high-pressure XRD, Raman spectroscopy, photoluminescence, and UV-Vis absorption measurements. High-pressure X-ray diffraction measurements on CaV$_2$O$_6$, reveal a pressure-dependent structural transition from the parent $C2/m$ phase to a lower symmetry $C2$ phase, with phase coexistence in the pressure range of 1 GPa to 5.2 GPa, accompanied by a change in coordination from VO$_5$ pyramids to VO$_6$ octahedra. High-pressure Raman measurements provide complementary evidence to the XRD study, with distinct changes in several Raman mode frequencies and their FWHM in the same pressure range. In addition, the optical measurements reveal the PL quenching and non-monotonic evolution of the optical band gap at around 1 GPa, where the transition from $C2/m$ to $C2$ phase starts. A similar pressure dependency of the optical band gap is observed in the sister compound MgV$_2$O$_6$, confirming that the optical band gap is highly dependent on coordination change and crystal symmetry.  Further, at higher pressure, the combined structural, vibrational, and optical study further reveals another isosymmetric transition in CaV$_2$O$_6$, without changing crystallographic symmetry.

\section{EXPERIMENTAL SECTION}
Polycrystalline CaV$_2$O$_6$ and MgV$_2$O$_6$ powder samples were prepared using the conventional solid-state reaction route. CaCO$_3$ and V$_2$O$_5$ were used as starting reagents for preparing CaV$_2$O$_6$ while MgO and V$_2$O$_5$ were used for MgV$_2$O$_6$. All the starting materials are procured from Sigma-Aldrich, with a 99.9\% purity. They were mixed in a stoichiometric ratio  and ground using an agate mortar and pestle. After grinding for 3 hours, the mixed powder was calcined in the furnace at about 550$^0$C for 5 hours. Next, the powder was pressed to make a pellet and sintered in air at about 650$^0$C for 5 hours to get the pure polycrystalline powder. The single monoclinic phase of the samples were confirmed by lab XRD.

High-pressure Raman, Photoluminescence, and UV-Vis absorption measurements were performed using a piston-cylinder type diamond anvil cell (DAC) with a culet size of 300 $\mu$m. A steel gasket of thickness 290 $\mu$m was preindented to a thickness of 40 $\mu$m. An electric discharge machine was used to drill a hole of diameter 100 $\mu$m at the center of the indented region of the gasket. A small amount of sample, along with pressure transmitting medium (PTM) and pressure calibrant were loaded into the central hole of the gasket. 4:1 methanol-ethanol mixture was used as PTM and Ruby fluorescence technique was used for pressure calibration.\\
High-pressure Raman spectroscopy measurements were carried out using a confocal micro-Raman spectrometer (Monovista from S\&I GmbH) in backscattering geometry, using the cobolt Samba 532 nm diode-pumped laser as excitation source. An infinity-corrected 20$\times$ objective with long working distance and focus spot size $\sim$ 5 $\mu$m was used to focus the laser beam and collect the backscattered Raman signal. The collected scattered light was dispersed using a grating with 1500 grooves/mm having a spectral resolution of about 1.2 cm$^{-1}$. 
High-pressure PL measurements were carried out using a confocal micro-Raman spectrometer (Horiba Jobin-Yvon LabRam HR-800) in backscattering geometry, equipped with 488 nm laser source. A grating of 600 grooves/mm was used to disperse the scattered light.
 High-pressure absorption measurements were performed using the Sciencetech 15189RD custom-made absorption spectrometer equipped with a broadband xenon light source LH-S-X300. To focus broadband light on the sample, an achromatic lens with a focal length of 19mm was used. The transmitted light was collected through a 10X objective and recorded using a Silicon detector.
High-pressure XRD measurements were performed at the PETRA III, P02.2 beamline at the wavelength $\sim$ 0.29 \AA \ and with XRD beam spot size (8$\times$3) $\mu$m$^2$. A small amount of sample, along with fine silver powder, were mixed well. The known equation of state of Ag was used for pressure calibration. 4:1 methanol-ethanol mixture was used as pressure transmitting medium. The sample-to-detector distance was calibrated by using the XRD pattern of standard material CeO$_2$. Dioptas software was used to integrate all the 2D diffracted image into  2$\theta$ vs intensity profiles \cite{prescher2015dioptas}. The Lebail and Rietveld refinements were performed using GSAS software \cite{expgui10graphical}. For analyzing XRD data, PCW \cite{kraus1996powder}, EOSFit7gui \cite{gonzalez2016eosfit7}, and VESTA \cite{momma2008vesta} software were used.

\section{RESULTS AND DISCUSSION}

	\subsection{High Pressure X-Ray Diffraction and Iso-symmetrical transition}
\begin{figure}[ht!]
	\centering
	\includegraphics[width=1.12\linewidth]{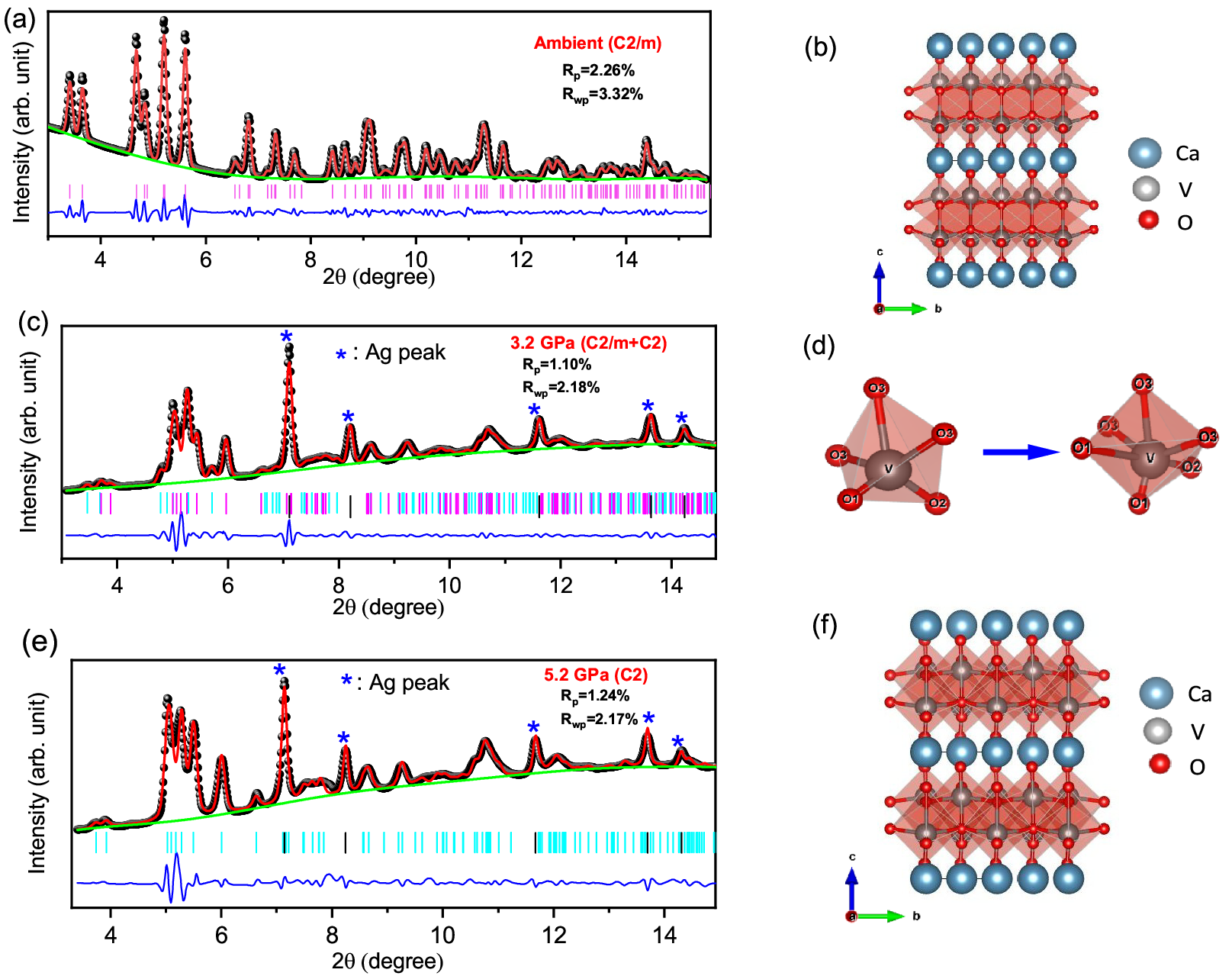}
	\caption{Rietveld refinement of synchrotron XRD pattern at (a)ambient pressure, (c) 3.2 GPa and (e) 5.2 GPa. The black circles show the experimental XRD data, while the red solid line corresponds to the Rietveld refinement fit to experimental data. The green line denotes the fitted background, and the blue line represents the difference between the experimental and calculated profiles. The magenta, cyan vertican lines indicate the Bragg reflection positions of $C2/m$ phase and $C2$ phase respectively. The Bragg reflection line of silver peak is marked by blue star. Crystal structure of CaV$_2$O$_6$, viewed along bc plane at (b)ambient pressure and (f) 5.2 GPa. The blue, grey and red spheres denotes Ca, V and O atoms respectively. (d) VO$_5$ bipyramid transformed to VO$_6$ octahedra.}
	\label{XRD Spectra}
\end{figure}

 At ambient conditions, all the diffraction peaks of CaV$_2$O$_6$ are indexed to the monoclinic crystal pseudo-Brannerite structure with space group symmetry $C2/m$. The Rietveld refinement of the ambient XRD pattern is carried out using the initial atomic positions taken from Yu et al. \cite{yu2015structure}. The Rietveld refined ambient XRD pattern 
  is shown in Fig.~\ref{XRD Spectra}(a). The refined structural parameters are $a = 10.063(7)$\AA, $b = 3.6743(2)$\AA, $c = 7.0403(5)$\AA, $\beta= 104.808(4)^0$ and unit cell volume $V = 251.68(3)$\AA$^3$ are in a good agreement with the previous literature \cite{islam2020comparative}. Table I represents the relative atomic positions after Rietveld refinement. The arrangement of CaV$_2$O$_6$ along the bc plane at ambient conditions is shown in Fig.~\ref{XRD Spectra}(b). CaV$_2$O$_6$ possesses a pseudo-Brannerite type structure where the distorted trigonal bipyramid VO$_5$ are connected by a zigzag chain along the b axis and form a layered structure along [001] direction, and the Ca atoms occupy the interlayer region forming CaO$_6$ octahedra.

\begin{table}[ht!] 
\caption{Relative atomic positions of CaV$_2$O$_6$ at ambient pressure for $C2/m$ phase}
	\begin{center}
		\begin{tabular}{ |c|c|c|c|c|c| } 
			\hline
			Atom & Wyckoff position & x/a & y/b & z/c\\
			\hline
			Ca & 2a & 0.0 & 0.0 & 0.0\\
			V & 4i & 0.2268(3) & 0 & 0.6861(4)\\
			O(1) & 4i & 0.0874 & 0.0 & 0.7219\\
			O(2) & 4i & 0.1434 & 0.5 & 0.1182\\
			O(3) & 4i & 0.2465 & 0 & 0.3934\\
			\hline
		\end{tabular}\\
	\end{center}
	\label{table:1}
\end{table}

\begin{figure}[ht!]
	\centering
	\includegraphics[width=1.1\linewidth]{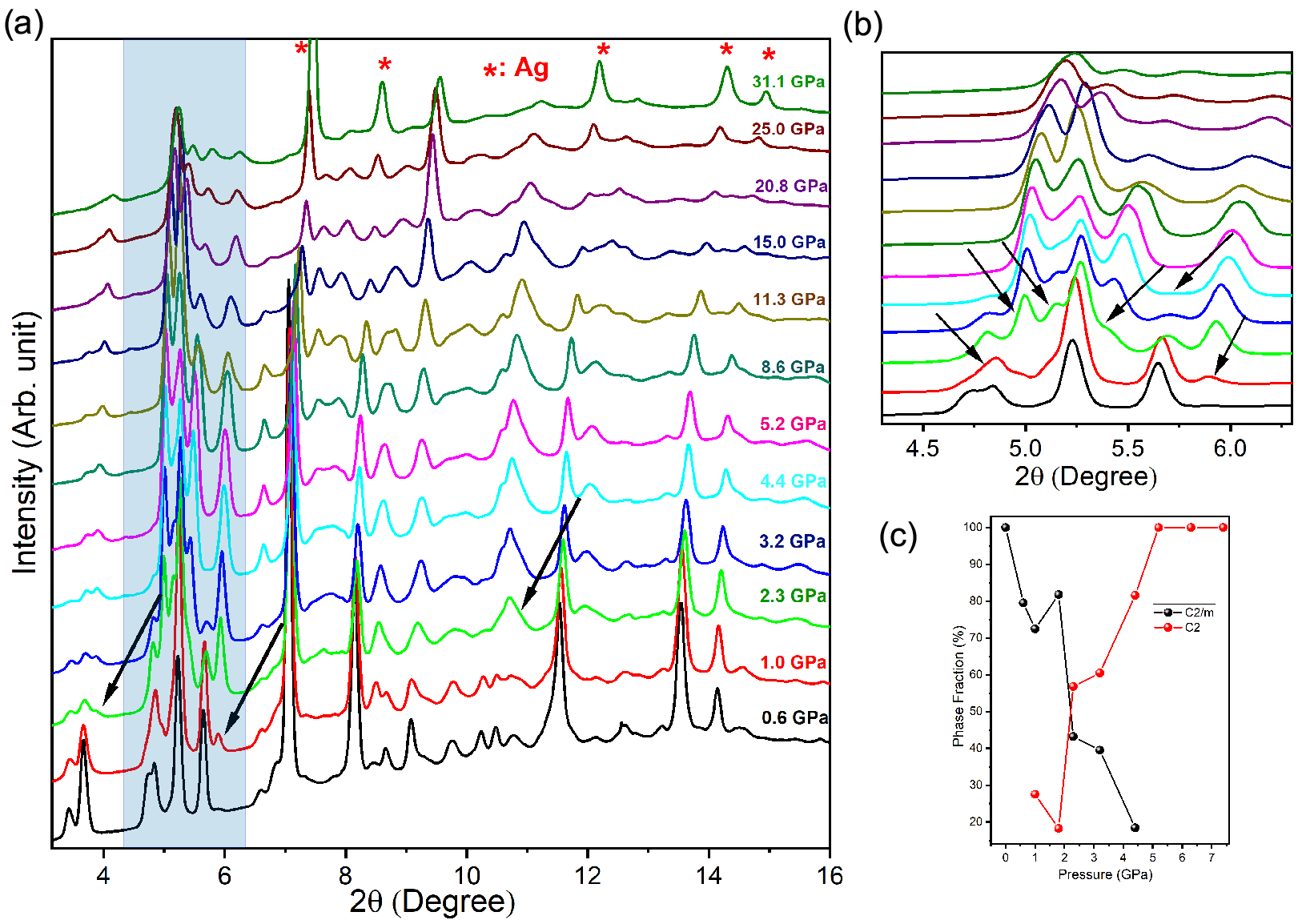}
	\caption{(a) XRD pattern of CaV$_2$O$_6$ at some selected pressures. The pressure values are indicated alongside the respective patterns. Black arrows mark the appearance of new diffraction peak. Red star indicates silver powder mixed with sample as a pressure marker. (b)Magnified view of the selected blue shaded region shown in (a). (c) Phase fraction at different pressure points for both $C2/m$ and $C2$ phases.}
	\label{HP XRD}
\end{figure}

High-pressure XRD measurements of CaV$_2$O$_6$ were carried out at room temperature up to 31.1 GPa. The pressure evolution of XRD patterns at selected pressure values is shown in Fig.~\ref{HP XRD}(a). Upon the application of external pressure, the system exhibits pronounced structural rearrangements. The Rietveld refinement is successfully carried out up to 0.6 GPa with a single monoclinic $C2/m$ phase. The onset of a structural phase transition occurs at approximately 1 GPa, marked by the emergence of a new Bragg reflection at 2$\theta$=5.89$^0$. As pressure increases, additional diffraction peaks emerge. The parent $C2/m$ phase alone failed to index the diffraction patterns. Careful observation shows a new diffraction pattern at 5.2 GPa. The pattern at 5.2 GPa could only be indexed to a $C2$ phase. This gives a wide structural phase coexistence window in the range 1 to 4.4 GPa.
Therefore, from 1 GPa to 4.4 GPa, we consider a two-phase model comprising $(C2/m+ C2)$, which results in an excellent fit to the experimental data. The Rietveld refinement of the XRD pattern at 3.2 GPa for the mixed $(C2/m + C2)$ phase (R$_p=1.10\%$ and R$_{wp}=2.18\%$) and at 5.2 GPa in the single $C2$ phase (R$_p=1.24\%$ and R$_{wp}=2.17\%$) are presented in Fig.~\ref{XRD Spectra}(c) \& (e). This structural transition results in a transformation of the local coordination of the V environment. In the parent $C2/m$ phase, V resides in a distorted VO$_5$ trigonal bipyramid.   
In the $C2$ phase, the coordination number of V increases and the VO$_5$ bipyramid transforms into distorted VO$_6$ octahedra, as depicted in Fig.~\ref{XRD Spectra}(d). These octahedra are interconnected by edge sharing, forming chains along the ab plane. Fig.~\ref{XRD Spectra}(f) represents the octahedral arrangements along bc plane at 5.2 GPa, as described above. The structure remains in the monoclinic $C2$ phase up to the highest pressure studied. Table II provides the relative atomic positions at 5.2 GPa for the $C2$ phase.

\begin{table}[ht!] 
\caption{Relative atomic positions of CaV$_2$O$_6$ at 5.2 GPa for $C2$ phase}
	\begin{center}
		\begin{tabular}{ |c|c|c|c|c|c| } 
			\hline
			Atom & Wyckoff position & x/a & y/b & z/c\\
			\hline
			Ca & 2a & 0.0 & 0.0095(12) & 0.0\\
			V & 4c & 0.1874(6) & 0.0079(14) & 0.7215(7)\\
			O(1) & 4c & 0.0875(17) & 0.0079(24) & 0.7219(21)\\
			O(2) & 4c & 0.1432(14) & 0.4911(21) & 0.1186(22)\\
			O(3) & 4c & 0.2467(16) & 0.0099(19) & 0.3932(19)\\
			\hline
		\end{tabular}\\
	\end{center}
	\label{table:2}
\end{table}

\begin{figure}[ht!]
		\centering
		\includegraphics[width=1.0\linewidth]{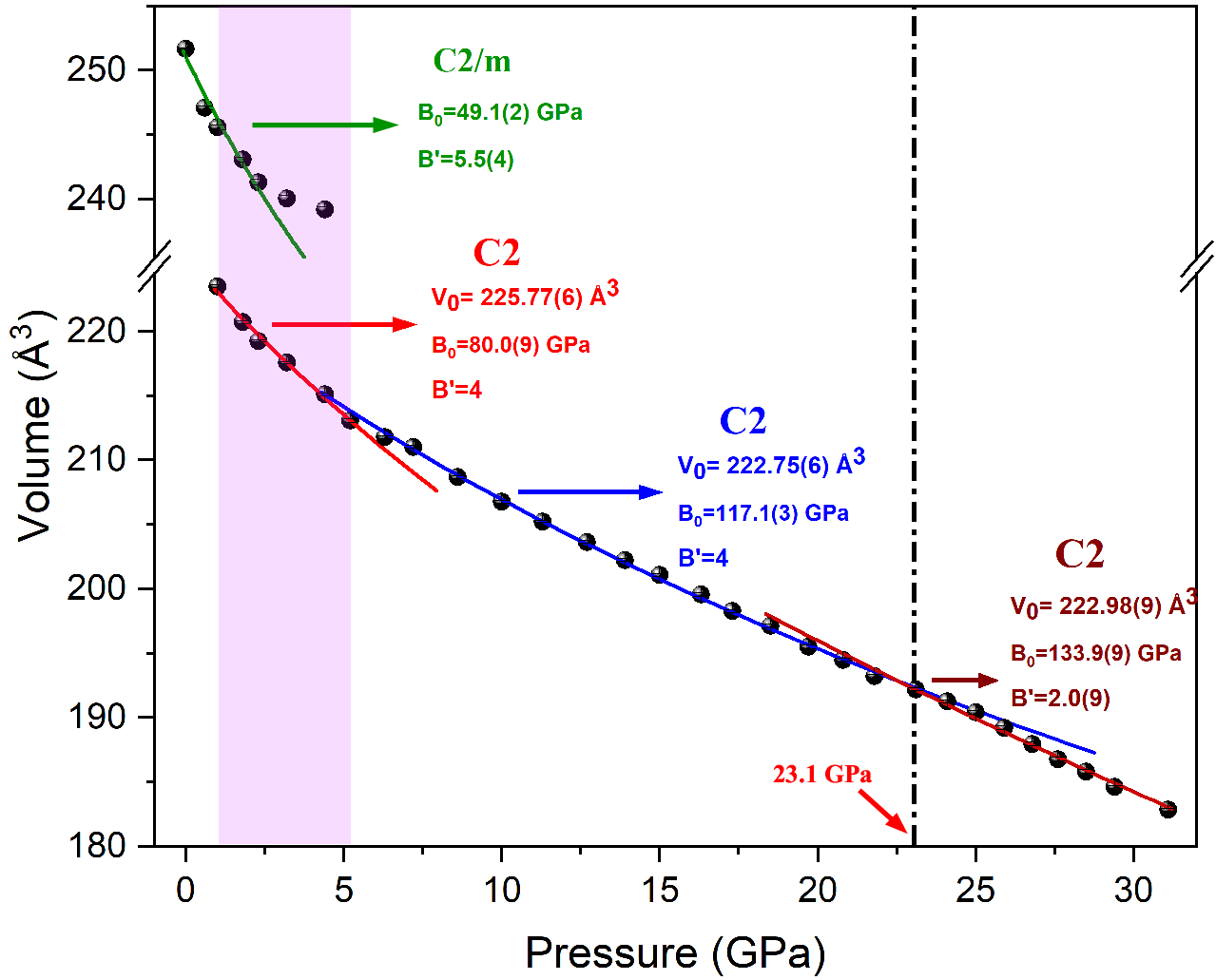}
		\caption{The evolution of unit cell volume of CaV$_2$O$_6$ with pressure. The green line shows Birch-Murnaghan(BM) EOS fit for $C2/m$ phase. The red and blue line indicate 2nd order BM EOS fit for $C2$ phase up to 23.1 GPa while the maroon line represents 3rd order BM EOS fit above 23.1 GPa to highest pressure studied. The purple shaded region indicates the mixed phase region where the C2/m and C2 phase coexist.}
		\label{EOSfit}
	\end{figure}

 The pressure evolution of unit cell volume is shown in Fig.~\ref{EOSfit} for both phases. One can see that the $C2$ phase results in a volume collapse. 
 The experimental volume data of both $C2/m$ and $C2$ phases are fitted with the Birch-Murnaghan equation of state (B-M EOS) \cite{ahmad2012theoretical}. 
 \begin{equation}
 	P(V)= \frac{3B_0}{2}\left[(\frac{V_0}{V})^{7/3}-(\frac{V_0}{V})^{5/3}\right]
 \end{equation}
 \begin{equation}
 	P(V)= \frac{3B_0}{2}\left[(\frac{V_0}{V})^{7/3}-(\frac{V_0}{V})^{5/3}\right]\times\left\{1+\frac{3}{4}(B^\prime-4)\left[(\frac{V_0}{V})^{2/3}-1\right]\right \}
 \end{equation}
 Equations 1 and 2 represent the 2nd- and 3rd-order B-M EOS, respectively, where $B_0$ is the isothermal bulk modulus, and $B^\prime$ is the first-order pressure derivative of the bulk modulus, and V$_0$ is the volume at the ambient pressure.

 For the $C2/m$ phase, the volume vs. pressure data is fitted with a 3rd-order B-M EOS, yielding a bulk modulus $B_0$=49.1 GPa and its first order derivative with respect to pressure is $B^{'}$=5.5. For the $C2$ phase, the complete volume dataset from 1 GPa to 31.1 GPa could not be fitted properly to a single equation of state. A careful examination reveals visible anomalies in the pressure evolution of volume data at about 4.4 GPa and at about 23.1 GPa. Therefore, we divide it into three pressure regions: the first region from 1 GPa to 4.4 GPa corresponds to the mixed phase region; the second region from 5.2 GPa to 23.1 GPa corresponds to the single phase $C2$ region, and both the above regions are fitted separately to the 2nd-order B-M EOS; and the third region from 23.1 GPa to 31.1 GPa is fitted with the 3rd-order B-M EOS. The first region fit results in a bulk modulus value of 80.0(9) GPa, substantially higher than the parent $C2/m$ phase. The second region and the third region result in bulk modulus values of 117.1(3) GPa and 133.9(9) GPa, respectively, with the $B^{'}$ value of 2.0(9) in the third region. The large increase in bulk modulus in the pure $C2$ phase indicates that the sample becomes less compressible at higher pressures. However, although the $B_0$ is slightly larger in the 3rd region compared to the 2nd region, the $B^{'} < 4$ means that the term ($B^{'}-4$) becomes negative in Eq.2, resulting in a softening of the rate at which the material becomes less compressible in the 3rd region. This indicates that instead pressure induced uniform shortening of bonds, which resists further compression (which should give $B^{'} > 4$), the unit cell framework of the crystal accommodates the extreme pressure by bending, tilting, or altering the internal symmetry of VO$_6$ and CaO$_6$ polyhedra resulting in an decrease in the rate of compressibility. This sudden change in compression behaviour occurs without any group-subgroup crystallographic symmetry changes. Therefore, one can describe it as an isostructural phase transition near 21–23 GPa \cite{ghosh2021strain,clarke2020isosymmetric}.
 
For a better understanding of the microscopic origin of the structural evolution under pressure, we have plotted pressure dependent lattice parameters for both phase $C2/m$ and $C2$ in Fig. S1. In the $C2/m$ phase, the lattice parameters decrease monotonically up to 4.4 GPa without any significant change, whereas the monoclinic angle $\beta$ increases up to 1 GPa, indicating the onset of a mixed $(C2/m+C2)$ phase. With further increase in pressure, the system gradually moves from the mixed phase toward a single $C2$ phase, which is evident from the variation of phase fraction under pressure as shown in Fig.~\ref{HP XRD}(c). It shows that the phase fraction of the $C2/m$ phase gradually decreases under pressure, while that of the $C2$ phase increases and reaches its maximum at 5.2 GPa. For the $C2$ phase, all the lattice parameters show anomalies around 5.2 GPa. The variation of angle $\beta$ in the $C2$ phase is shown in Fig. S3(a). The relative lattice parameters as a function of pressure for both phases are presented in Fig. S3 (b) and (c) respectively. For the $C2/m$ phase,  the lattice parameters undergo an anisotropic compression under pressure. As the chain of VO$_5$ bipyramids forms a layer along the c axis, the relatively weak interlayer interactions result in enhanced compressibility along the c axis compared to the a and b axes. In contrast, for the $C2$ phase, all the lattice parameters show isotropic compression up to 5.2 GPa. With further compression, the a axis becomes more compressible as compared to the c axis. In the $C2$ phase, VO$_6$ octahedra form chains within the ab plane. This rearrangement of VO$_6$ octahedra in the $C2$ phase possibly gives rise to contrasting anisotropic compression behaviour between $C2/m$ and $C2$ phases.

\begin{figure}[ht!]
	\centering
	\includegraphics[width=1.1\linewidth]{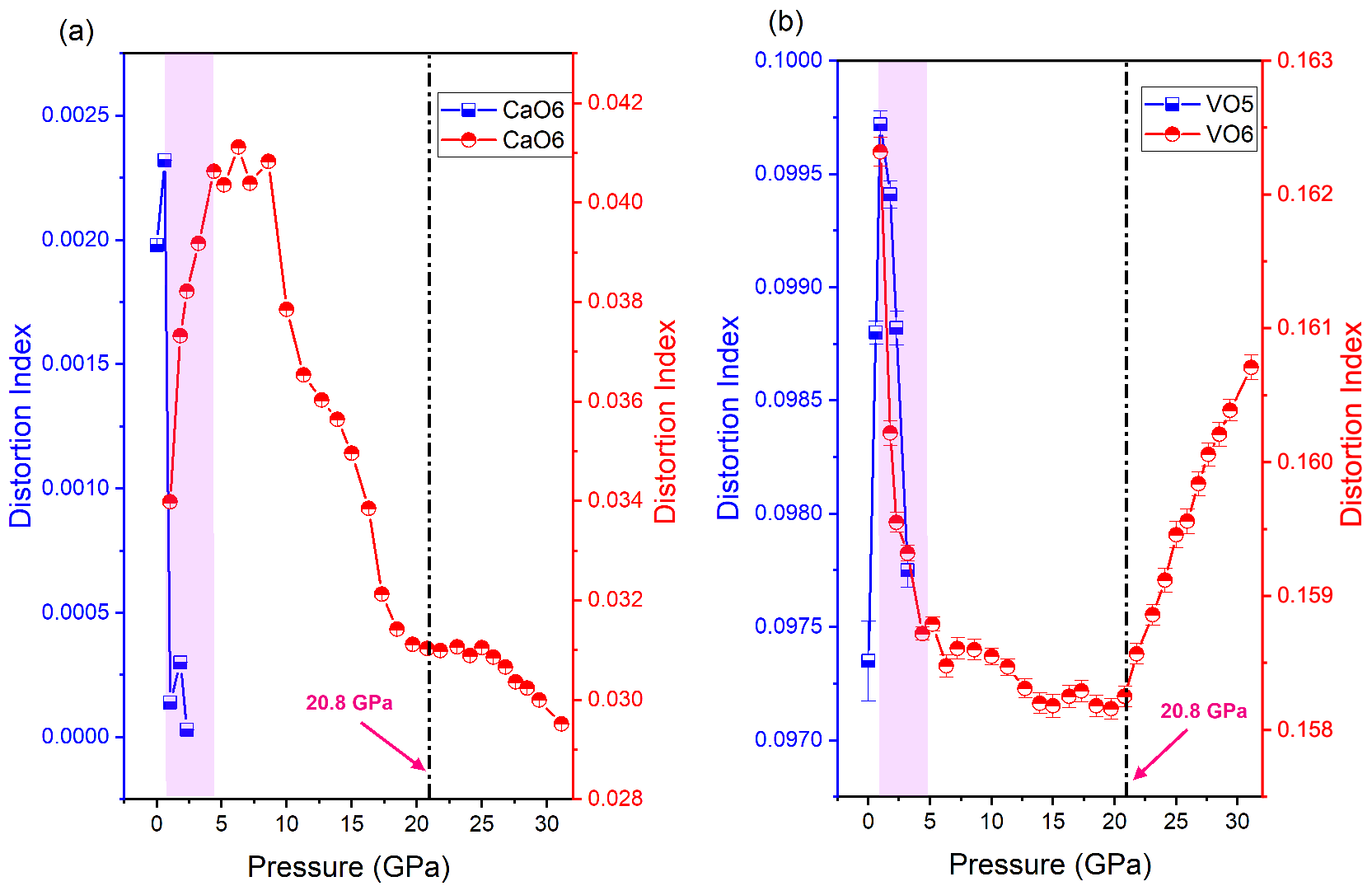}
	\caption{(a) The pressure evolution of distortion index for CaO$_6$ octahedra for both $C2/m$ and $C2$ phases. (b) Pressure evolution of distortion index of VO$_5$ pyramid for $C2/m$ phase and VO$_6$ octahedra for $C2$ phase. Blue and Red symbols correspond to $C2/m$ and $C2$ phases respectively. Purple shade indicates the mixed $(C2/m+C2)$ phase region. The black dashed line marks anomaly at 20.8 GPa.}
	\label{distotion index}
\end{figure}

To understand the anomalous compression behavior in more depth, we have investigated the effect of pressure on different polyhedra for both phases and shown in Fig.~\ref{distotion index}(a) \& (b). The distortion index of a polyhedron can be defined as,
 \begin{equation}
     DI= \frac{1}{N}\sum_{i=1}^{N}\frac{l_i-l_{av}}{l_{av}}
\end{equation}
where N denotes the number of bonds of a polyhedron, $l_i$ is the individual bond length, and $l_{av}$ is the average bond length.
 For the $C2/m$ phase, the distortion index of the CaO$_6$ octahedra decreases with increasing pressure and becomes almost constant in the mixed phase region. In contrast, the distortion of the VO$_5$ pyramid initially increases up to 1 GPa and subsequently decreases in the mixed phase region where the contribution of the $C2/m$ phase gradually decreases, as shown in Fig.~\ref{HP XRD}(c). In the mixed phase region, the contribution of the $C2$ phase gradually becomes dominant with pressure, accompanied by a reduction in the distortion index of VO$_6$ octahedra, which indicates a tendency to form more regular octahedra. However, with increasing pressure, CaO$_6$ octahedra exhibit the opposite behaviour compared to VO$_6$.  Above 5.2 GPa, a gradual decrease in distortion is observed for CaO$_6$ octahedra, indicating a gradual approach to a regular octahedron, while the VO$_6$ octahedra show nearly pressure independent distortion due to the compact structure along the ab plane. Interestingly, around 20.8 GPa, an increase in the distortion index is observed again for VO$_6$ octahedra. The enhanced distortion index of VO$_6$ octahedra contributes to the anomalous compressibility behaviour in this pressure range, which drives the isosymmetric phase transition as mentioned above.

\subsection{High Pressure Raman Spectroscopy and Lattice Anharmonicity}

\begin{figure}[ht!]
	\centering
	\includegraphics[width=1.1\linewidth]{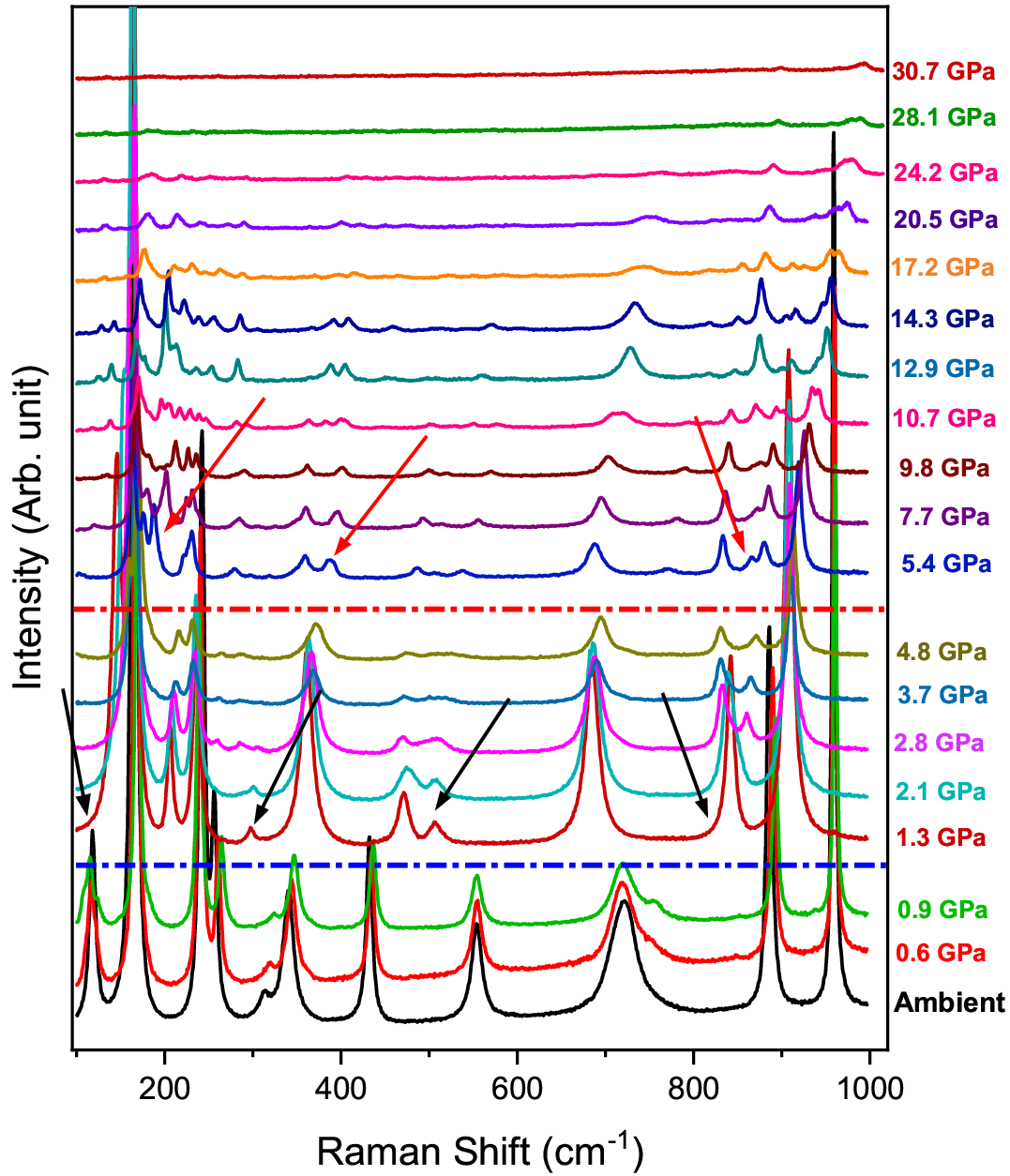}
	\caption{Pressure evolution of Raman spectra at some selected pressure points. Black and red arrows indicate appearance of new modes, as well as the splitting and merging of several existing modes. The horizontal blue and red dashed lines mark the transition pressure region at around 1 GPa and 5.2 GPa. }
	\label{HP Raman}
\end{figure}

 To decode the macroscopic mechanical transitions from a lattice dynamics perspective, we track the experimental vibrational modes by Raman spectroscopy measurements.
 High-pressure Raman spectroscopy probes pressure-induced changes in lattice vibrations, revealing structural distortions, phase transitions, and electron–phonon interactions under compression \cite{mishra2026pressure,sahu2025pressure,mukherjee2024pressure,mukherjee2025soft}. Group theoretical analysis shows that the $C2/m$ space group produces a total of 24 optical modes, and it is represented as (4A$_u$+8B$_u$+8A$_g$+4B$_g$), in which a total of 12 modes (8A$_g$+4B$_g$) are Raman active \cite{jin2022comprehensive}. At ambient conditions, our experiments reveal  11 Raman active modes, which are in good agreement with the literature \cite{jin2022comprehensive}. At each pressure point, all the Raman modes are fitted to a Lorentzian function. Fig. S5(a) shows the fitted Raman spectrum at ambient pressure. Each mode is labelled as M$_i$ (where i=1 to 11). The low-frequency modes M$_1$ (118 cm$^{-1}$) and M$_2$ (165 cm$^{-1}$) originate from the stretching vibration of V$_2$O$_2$ bridges, formed by edge sharing between two adjacent pyramids, whereas M$_3$ (242 cm$^{-1}$), M$_4$ (256 cm$^{-1}$), and M$_6$ (340 cm$^{-1}$) modes are attributed to the vibration of CaO$_6$ octahedra. The weak M$_5$ (313 cm$^{-1}$) mode is associated with the bending vibration of V$_3$O unit, where one oxygen atom is shared with three V atoms from three different VO$_5$ pyramids. The intermediate frequency modes M$_8$(553 cm$^{-1}$) and M$_9$ (720 cm$^{-1}$) are attributed to the combination of the stretching vibration of  V$_2$O$_2$ bridges and V$_3$O unit. The M$_7$ (431 cm$^{-1}$) and M$_{10}$ (885 cm$^{-1}$) modes are assigned to the bending vibration of V-O-V bridges (oriented parallel to the b axis of the unit cell) due to symmetric and antisymmetric vibrations, respectively. The high frequency mode M$_{11}$ (958 cm$^{-1}$) is attributed to the stretching vibration of the terminal V-O bond, which is characterized by a very strong V-O bond with the shortest bond length \cite{jin2022comprehensive,chen2004raman,baran1987raman}. The pressure evolution of Raman spectra at some selected pressures is shown in Fig.~\ref{HP Raman}. Around 1.3 GPa, we can see a pronounced rearrangement of Raman spectra, which probably indicates a structural transition to the mixed phase, as observed in the XRD study. We have reassigned the Raman modes in the mixed phase as N$_i$ (where i=1 to 10). For clarity, The Raman spectra at 0.9 GPa and 1.3 GPa are compared in Fig. S5(b), where the rearrangement of Raman spectra can be clearly identified. As the system approaches towards the $C2$ phase, several new Raman modes P$_1$ (258 cm$^{-1}$), P$_2$ (286 cm$^{-1}$), and P$_3$ (854 cm$^{-1}$) start appearing. With further increase in pressure, in the pure $C2$ phase, some of the Raman modes split and are marked by arrows  in Fig.~\ref{HP Raman}.

\begin{figure}[ht!]
	\centering
	\includegraphics[width=1.1\linewidth]{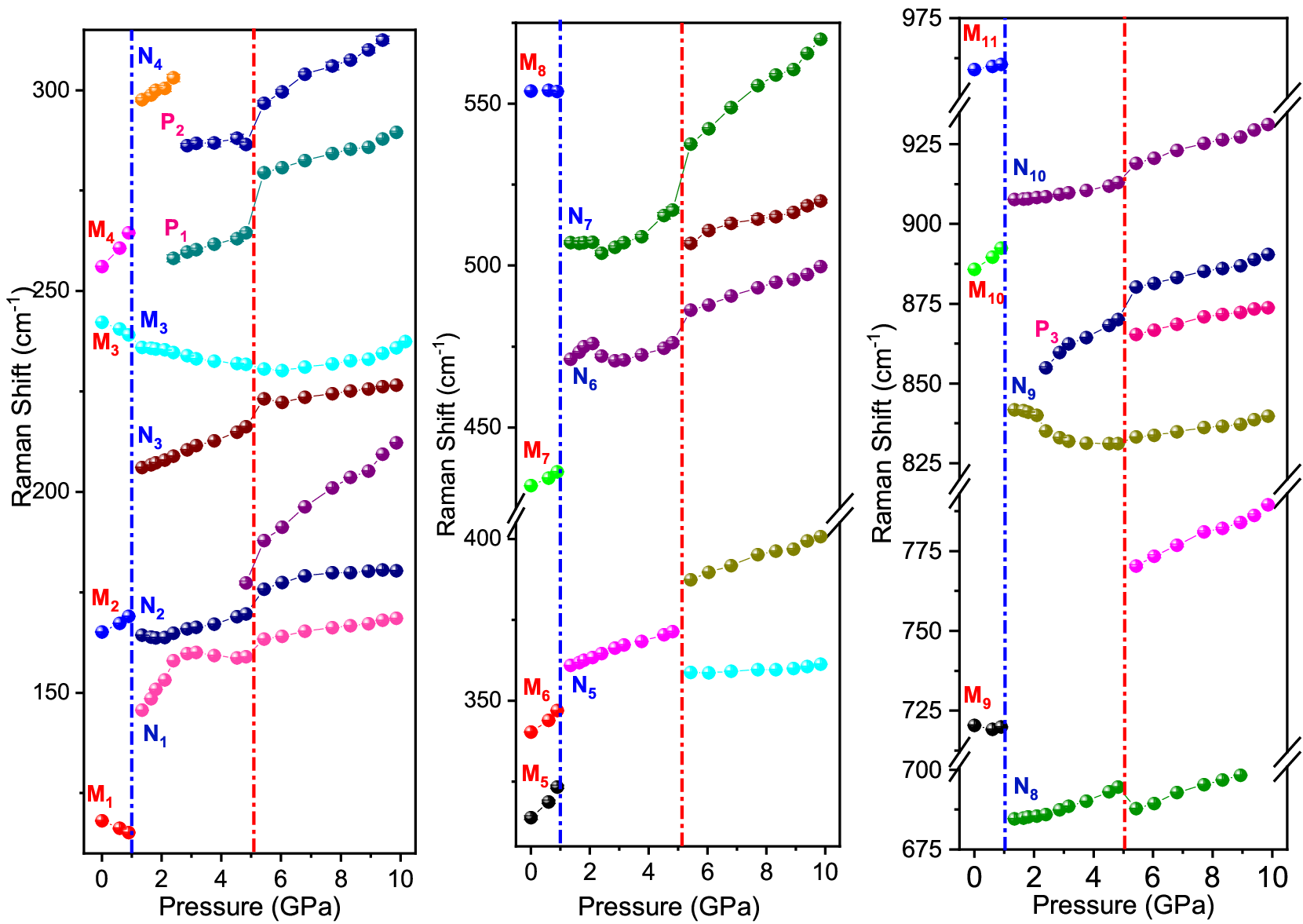}
	\caption{Pressure evolution of all the Raman modes of CaV$_2$O$_6$. The blue and red dashed line represent transition pressure region at 1 GPa and 5.2 GPa.}
	\label{All raman modes}
\end{figure}

The evolution of all the Raman peak positions under pressure is depicted in Fig.~\ref{All raman modes}. Most of the Raman modes exhibit a blue shift under pressure. The hardening of the Raman modes is associated with the shortening of the average bond length, depicted in Fig. S2. As the bond length decreases with pressure, the bond strength increases, resulting in the hardening of these Raman modes. Interestingly, in the $C2/m$ phase, the low frequency modes M$_1$ and M$_3$ show anomalous softening, with M$_1$ disappearing above 1 GPa. Additionally, in the mixed phase, the M$_3$ and N$_9$ modes keep softening till 5.2 GPa, which indicates instability in the system. To quantify the lattice anharmonicity and map the sublattice instabilities driving the structural transformation, we calculate the experimental isothermal Mode-Grüneisen parameters ($\gamma _{i}$) for the observed phonons using the following equation:
\begin{equation}
    \gamma_i = \frac{B_0}{w_0}\left(\frac{dw_i}{dP}\right),
\end{equation}
where $B_0$ is the isothermal bulk modulus, and $w_0$ is the ambient frequency.
The low-frequency modes $M_{1}$ (V$_2$O$_2$ bridge stretching) and $M_{3}$ (CaO$_{6}$ vibration) gives negative Grüneisen parameter values of $\gamma_1$ = -1.34 and $\gamma_3$ = -0.79, respectively. Similarly in the mixed phase region, the modes  M$_3$ and N$_9$ yields the values of $\gamma_3$ = -0.42 and $\gamma_9$ = -0.46. A negative Mode-Grüneisen parameter serves as a signature of optical phonon instability, and therefore represents a progressive reduction of the structural restoring force that ultimately drives the structural collapse into the $C2$ symmetry above 1 GPa. As the system crystallizes into the single stable $C2$ phase at 5.2 GPa, every single mode exhibits a discontinuity with a sharp inflection in its pressure coefficient $d\omega/dP$, mirroring the macroscopically observed jump in bulk modulus from 80.0 GPa to 117.1 GPa. Also, any changes in lattice instability can be reflected in the internal strain of the system. To understand the effect of external pressure on the internal strain we have plotted Reduced pressure vs. Eulerian strain plot, and discussed in the section A. of the supplementary material.

\begin{figure}[ht!]
	\centering
	\includegraphics[width=1.0\linewidth]{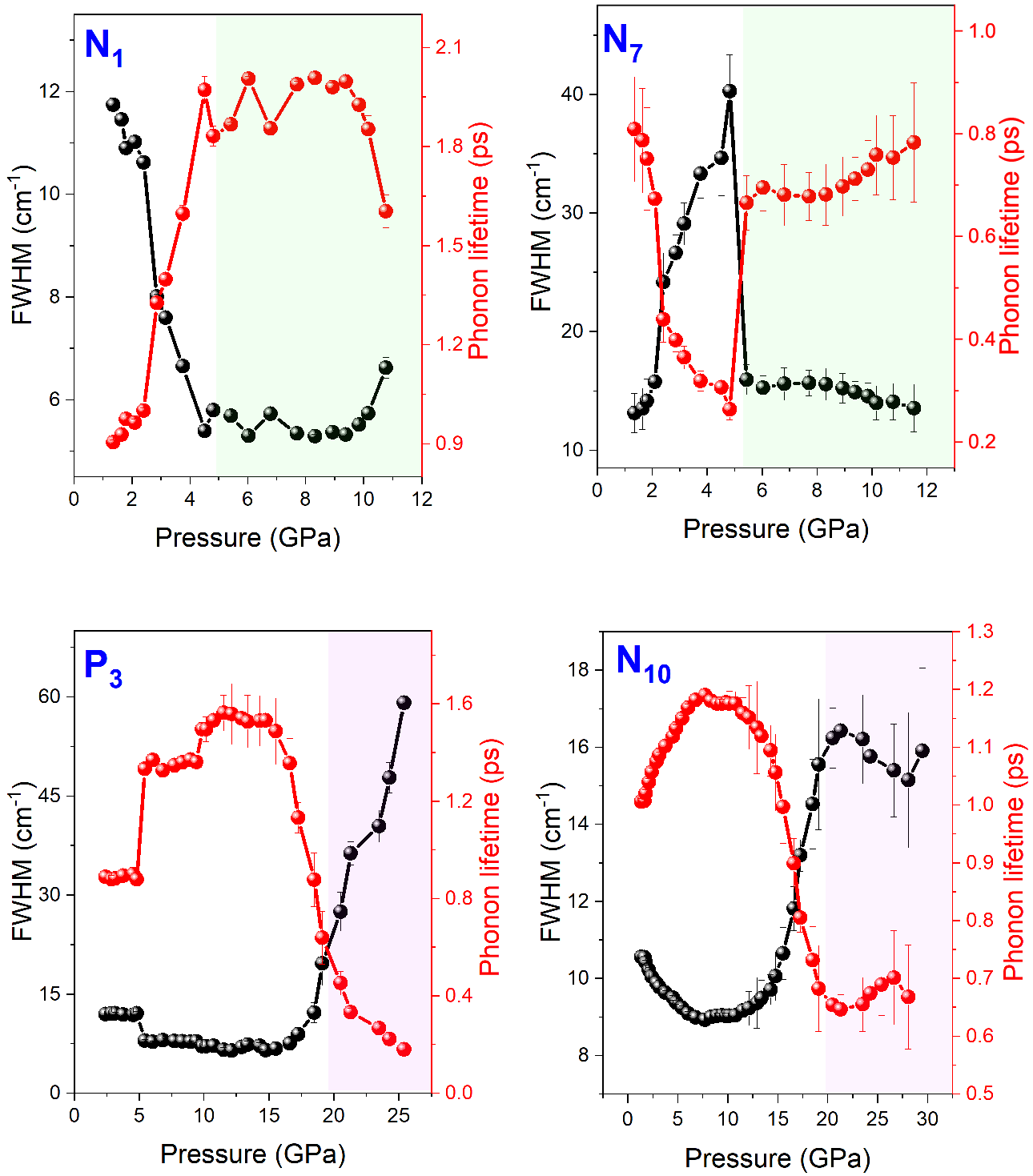}
	\caption{Pressure evolution of FWHM and corresponding phonon lifetime for some selected Raman modes of CaV$_2$O$_6$. Black and Red sphere correspond to FWHM and phonon lifetime respectively. .}
	\label{FWHM}
\end{figure}

 The contribution from anharmonic phonon-phonon interaction and electron-phonon coupling is reflected in the phonon lifetime, which is inversely proportional to the linewidth of a Raman mode given by its full-width-half-maximum (FWHM) \cite{bera2013sharp,rajaji2018pressure,samanta2025pressure}. 
 To probe how the transitions modify phonon-phonon scattering landscapes, the FWHM $(\Gamma)$ data is converted into experimental phonon lifetimes ($\tau $) using the energy-time uncertainty relation: $\tau =\frac{1}{\pi c\Gamma }$, where $c$ is the velocity of light and are presented for certain selected Raman modes in Fig.~\ref{FWHM}.
 In our study, we observe various anomalies in $\Gamma$ and hence in $\tau$. 
The N$_{1}$ mode linewidth shows anomalous narrowing up to 5.2 GPa, translating to a substantial increase in phonon lifetime $\tau$ that expands from $\sim$ 0.9 $ps$ to $\sim$ 2 $ps$. Similarly, an increase in the phonon lifetime is also observed for N$_7$ mode at 5.2 GPa. This reduction in scattering rates indicates that entering the pure $C2$ single-phase configuration cleanses the structural disorder of the intermediate mixed state, effectively quenching anharmonic phonon decay pathways.
Above 20 GPa, the high frequency N$_{10}$ and P$_3$ modes (localized inside the VO$_6$ octahedra) exhibit a sudden, sharp broadening anomaly. This sudden contraction of the phonon lifetime directly reflects enhanced scattering from the structural strain and local polyhedral distortions (Fig.~\ref{distotion index}) accompanying the isosymmetric transition. The absence of any volume discontinuity along with any drastic change in Raman spectra confirm this iso-symmetric transition, which is driven by only elastic anomalies.
 
The optical and electronic properties of the systems are highly dependent on the structural geometry. The transition from the centrosymmetric $C2/m$ phase to non-centrosymmetric $C2$ phase leads to a removal of inversion symmetry and associated selection rule for optical transitions. Additionally, pressure-induced changes in V-O coordination can alter the electronic state. Hence, a detailed high-pressure photoluminescence and absorption measurement are performed to understand the correlation between structural and optical properties.

\subsection{Optoelectronic Transitions and Microscopic Structure-
Property Relation}

The ambient PL spectrum is recorded using the 488 nm laser, which is shown in Fig. S6(a). It exhibits a long tail at lower energy, with a broad emission centered at around 705.81 nm, with FWHM of 176.28 nm. The absorbance spectrum at ambient condition is shown in Fig. S6(b), with absorption peak centered at around 423 nm with a Stokes shift of 1.15 eV. To determine the optical band gap of CaV$_2$O$_6$, we have employed Tauc's plot with $(\alpha h \nu)^n=A(h\nu-E_g)$, where $\alpha$ is absorption coefficient, A is a constant, $h\nu$ is the energy of photon, $E_g$ is the optical band gap of the material and n=2 and $\frac{1}{2}$ for direct and indirect band gap, respectively \cite{tauc1966optical,zhong2023idealizing}. Fig. S6(b) shows ambient absorption spectra fitted with Tauc's plot in the linear absorption edge for indirect band gap calculation. The band gap obtained from the calculation is 2.52 eV. Experimentally, from absorption spectroscopy, the value of band gap reported in CaV$_2$O$_6$ ranges from around 2.56 eV to 2.66 eV \cite{yu2015structure,jin2022comprehensive,islam2020comparative}. The slight differences in values can be attributed to Urbach tail in the experimental Tauc's plot \cite{garg2023accurate}. But our measured value is close to the reported literature. The steep absorption edge indicates that the observed optical absorption is primarily associated with fundamental band-to-band transitions rather than transitions involving impurity or defect states. The PL spectrum exhibits a luminescence peak at 1.75 eV, whereas the optical band gap obtained from UV-Vis absorption spectra is 2.52 eV, which suggests that the origin of the PL is not associated with the band to band recombination. According to Jin et al. PL emission is attributed to defect states such as oxygen vacancy or donor-acceptor pairs within the sample \cite{jin2022comprehensive}.  The observed broad emission with a large Stokes shift can be related to the lattice distortion resulting in defect states, giving rise to strong electron-phonon coupling.  

\begin{figure}[ht!]
	\centering
	\includegraphics[width=1.0\linewidth]{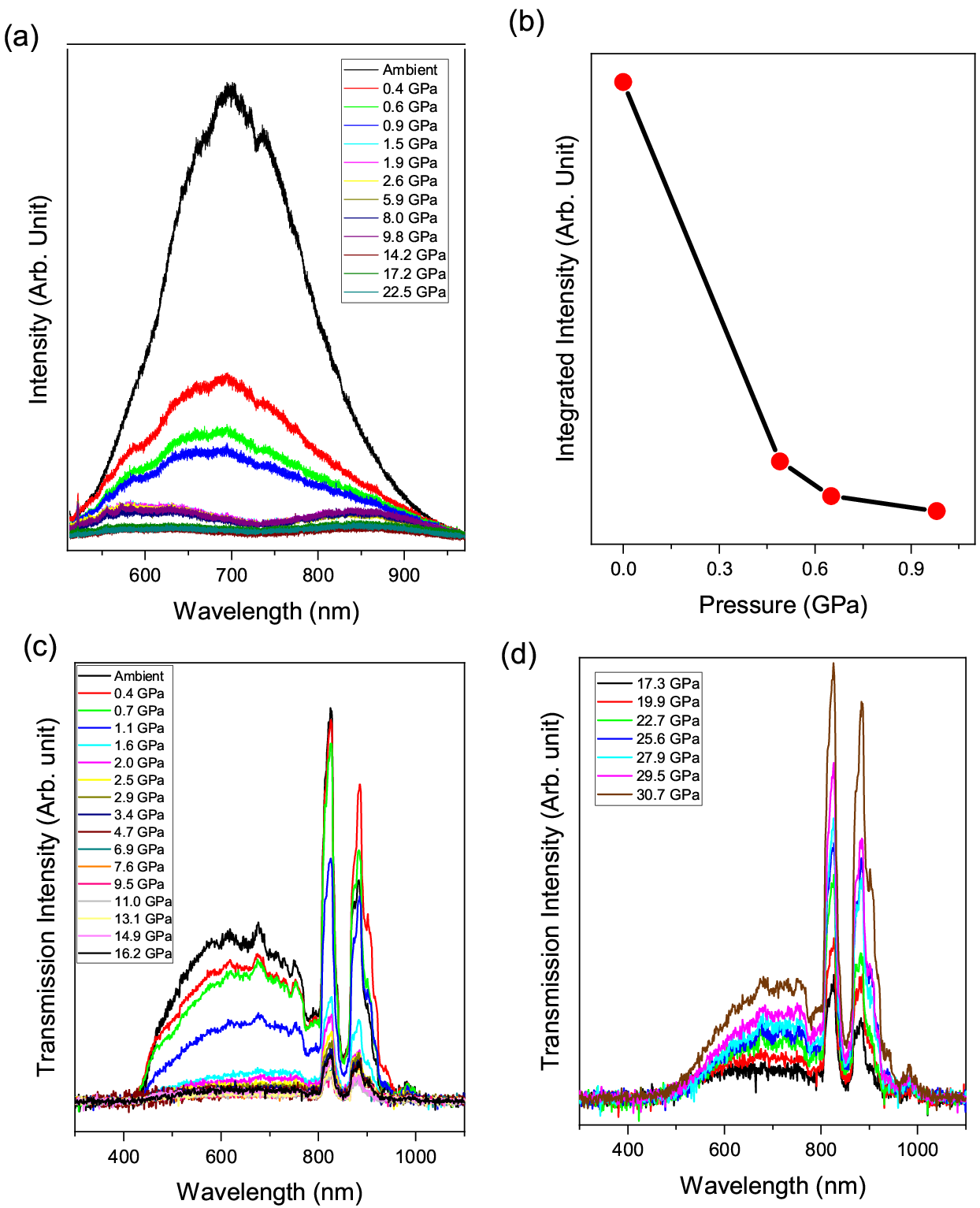}
	\caption{(a) Evolution of PL spectra of CaV$_2$O$_6$ at some selected pressures. (b) Variation of integral intensity with pressure. (c) \& (d)Transmission spectra of CaV$_2$O$_6$ at some selected pressure points, measured using a xenon source .}
	\label{PL Transmission}
\end{figure}

The pressure evolution of PL spectra of CaV$_2$O$_6$ is shown in Fig.~\ref{PL Transmission}(a). A significant quenching of PL intensity is observed with increasing pressure, as shown in Fig.~\ref{PL Transmission}(b), and the PL intensity quenches around 1 GPa. One can immediately see that this optical quenching is directly related to VO$_5$ to VO$_6$ coordination change.
Since the PL originate from defect states, the photoexcited carriers are trapped in defect states and radiatively recombine, which produces PL in the system. The PL quantum efficiency can be expressed as 
\begin{equation}
    \eta_{PL}= \frac{K_r}{K_r + K_{nr}}
\end{equation}
where $K_r$ and $K_{nr}$ are the radiative and non-radiative recombination rates, respectively. Therefore, if $K_{nr}$ increases with pressure, then subsequently intensity of PL decreases. With increase in pressure, the creation of intermediate, highly distorted coordination configurations drastically raises the localized defect-state density. This opens up non-radiative relaxation pathways, where photoexcited carriers transfer their energy to lattice vibrations (phonons) rather than radiative photons leading to PL quenching.

\begin{figure}[ht!]
	\centering
	\includegraphics[width=1.0\linewidth]{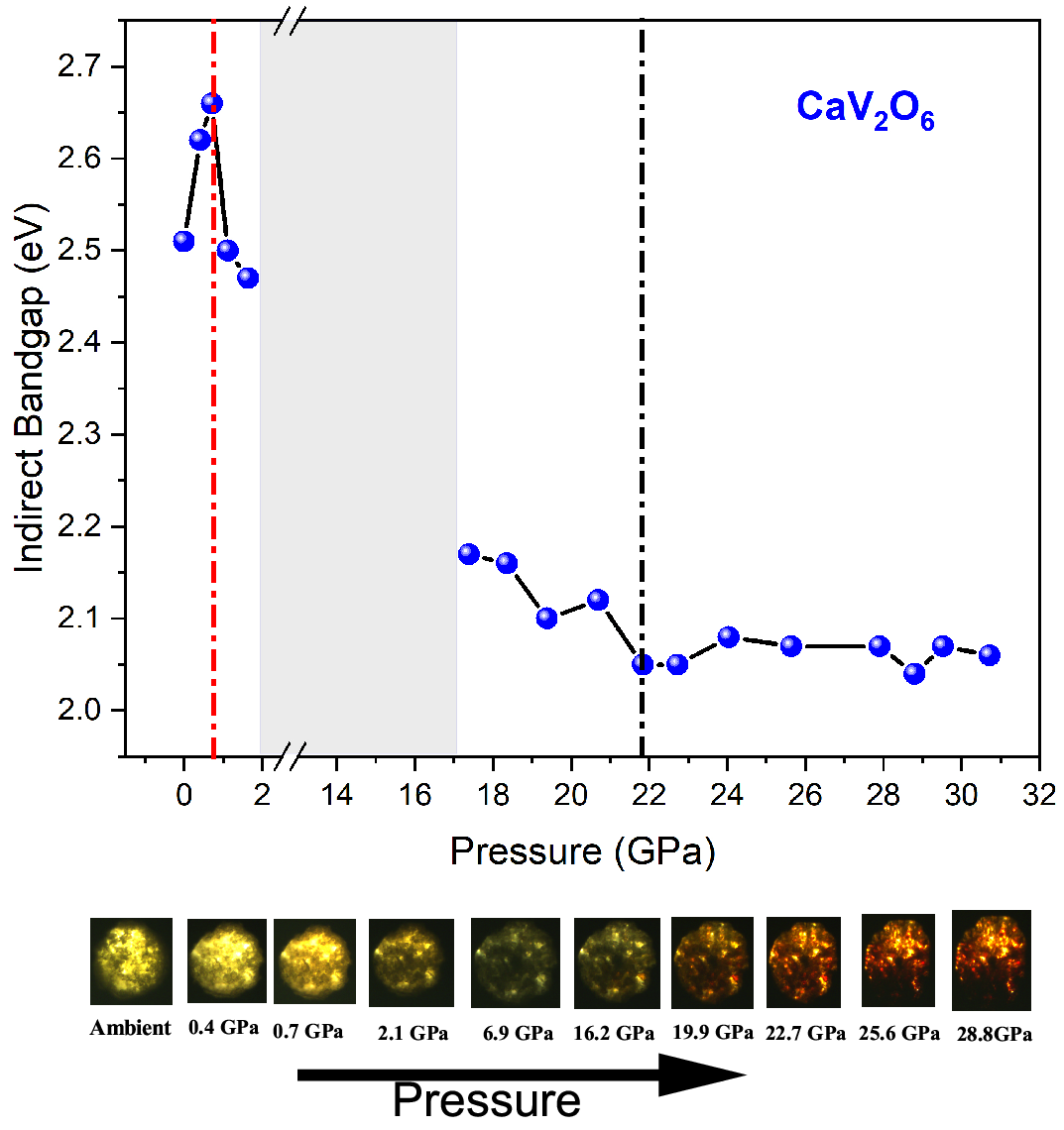}
	\caption{Evolution of Indirect Band gap value of CaV$_2$O$_6$ at different pressure. The black shaded region represents the pressure range where reliable measurement of optical band gap is not possible due to strong absorption. The red and black dashed line correspond to pressure points around 1 GPa and 22 GP. The pressure induced color changes of the sample are shown below.  }
	\label{Bandgap}
\end{figure}

The transmission spectra of CaV$_2$O$_6$, obtained from the UV-Vis absorption measurement at various pressure points is plotted in Fig.~\ref{PL Transmission} (c) and (d). A remarkable decrease in transmission intensity is observed with increasing pressure. Around 1.6 GPa, the intensity is reduced to approximately 70\% of the ambient value, which indicates the increase of absorption of light by the sample. The pressure-dependent band gap of the material is calculated using Tauc's plot, shown in Fig.~\ref{Bandgap}. With the increase in pressure, the optical band gap increases to about 2.67 eV at about 1 GPa, exactly tracking the pressure range where the distortion index of the parent
VO$_5$ bipyramid increases and the system enters into the mixed phase $(C2/m+C2)$ region. Further increase in pressure results in a decrease of the band gap. As discussed above, the optical absorption is primarily associated with fundamental band to band transitions. The pressure derivative of the optical band gap, $\frac{dE_g}{dP}>0$ below 1 GPa and $\frac{dE_g}{dP}<0$ above 1 GPa, which suggests that there is a reorganization of band edges. Zhao et al. \cite{zhao2024coordination} showed that for the metavanadate LiVO$_3$, at the structural transition pressure, VO$_4$ tetrahedra change coordination to VO$_6$ octahedra, resulting in an electron density rearrangement around the Li and O ions. Similarly, a high-pressure study on InTaO$_4$ reported an increase in the coordination of both polyhedra around the structural transition, accompanied by an initial increase in the band gap with subsequent decreases after the structural transition. In CaV$_2$O$_6$, the valence-band and conduction-band edges are predominantly dominated by the O-2p and V-3d orbitals, respectively \cite{jin2022comprehensive}. Therefore, the evolution from five coordinated VO$_5$ units toward edgesharing distorted VO$_6$ octahedra reorganizes the local ligand-field environment of V and can modify the pressure dependence of V-3d derived conduction states and their hybridization with O-2p states. This provides a plausible explanation for the change in the pressure coefficient of the optical gap near the structural transition region. Therefore, the onset of $C2/m$ to $C2$ transition provides a natural microscopic origin for the anomalous pressure dependence of the optical gap.

With further increase in pressure above 1 GPa, the transmission intensity progressively decreases, accompanied by a gradual color change of the sample from yellow to black, suggesting pressure-induced piezochromic behavior. The observed color change is associated with the O$^{2-}$ to V$^{5+}$ charge transfer mechanism  \cite{lakshminarasimhan2022optical}. In the present case, pressure-induced increase in coordination may influence the O$^{2-}$ to V$^{5+}$ charge transfer states. Such a modification in the electronic states can increase the optical absorption and subsequently change the color from yellow to black. Further increase in pressure results in high absorption, which makes the transmission signal intensity very low, preventing us from reliably calculating the band gap in the 2 GPa to 16 GPa pressure region. 

\begin{figure}[ht!]
	\centering
	\includegraphics[width=1.0\linewidth]{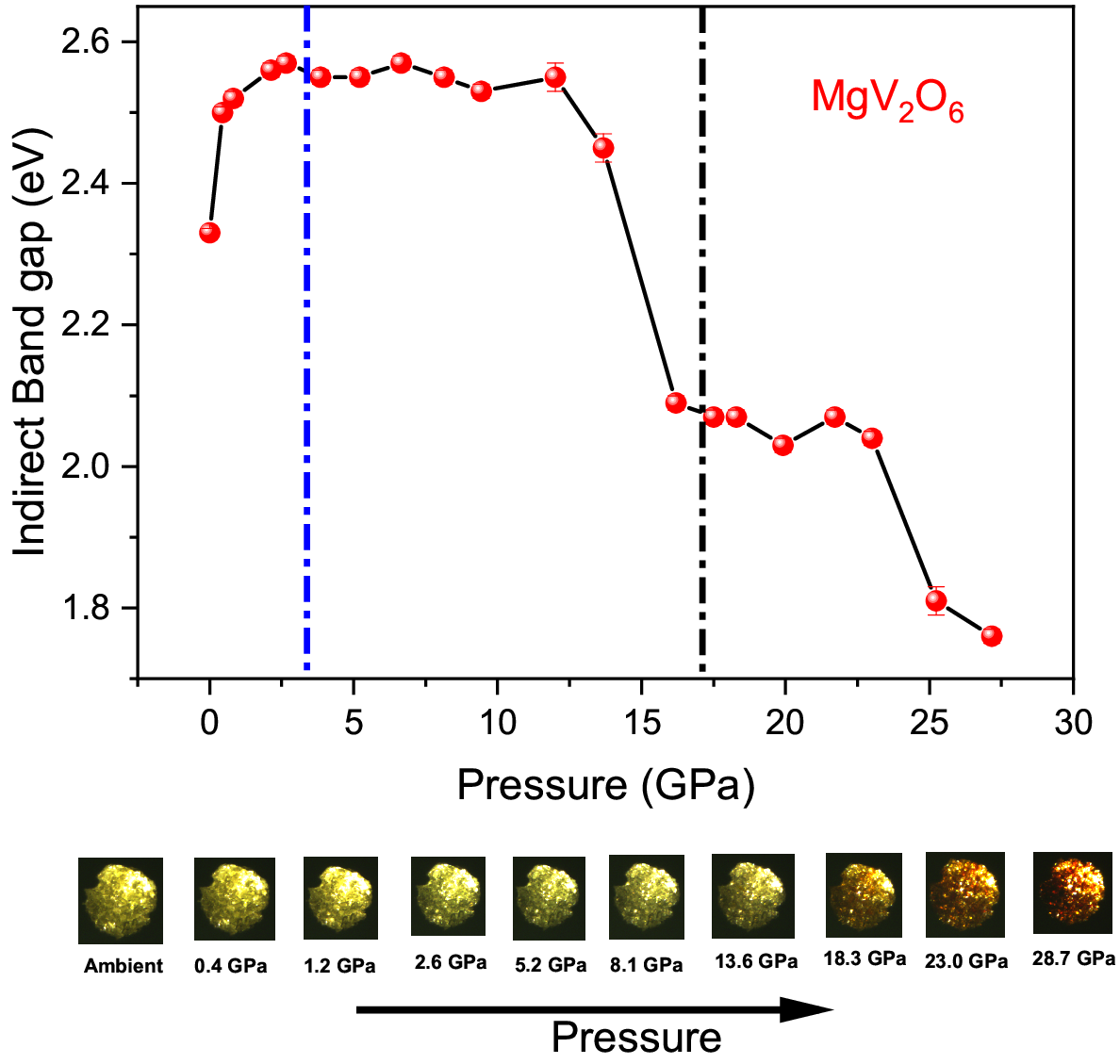}
	\caption{Evolution of Indirect Band gap value of MgV$_2$O$_6$ at different pressure. The blue and black dashed line marked around 4 GPa and 17 GPa.  The pressure induced color changes of the sample is shown below.  }
	\label{MVO Bandgap}
\end{figure}

This structural-optical scaling law is verified by our parallel high-pressure tracking of the sister compound MgV$_2$O$_6$. Mg$V_2$O$_6$ exhibits a pressure induced change in V coordination and undergoes a structural phase transition from single $C2/m$ phase to a mixed $C2/m+C2$ phase \cite{tang2016exploring}. To elucidate the effect of these structural changes on the optical properties and to compare its pressure response with CaV$_2$O$_6$ results, we have carried out a high-pressure optical absorption measurement on MgV$_2$O$_6$. We have also performed high-pressure Raman measurements and the corresponding results are shown in Fig. S7 of the supplementary material. The high-pressure Raman results are consistent with the already reported literature \cite{tang2016exploring} and exhibit distinct anomalies around 4 GPa and 17 GPa, where the sample undergoes a coordination change in V and a structural transition, respectively. It is important to mention that no high-pressure optical band gap measurements on MgV$_2$O$_6$ are available in the literature, and this study reports them here for the first time to the best of our knowledge. At ambient conditions, MgV$_2$O$_6$ is found to be an indirect band gap semiconductor, with an optical band gap of 2.3 eV, which matches well with the previous literature \cite{li2015effects}. The pressure evolution of the optical band gap of MgV$_2$O$_6$ is shown in Fig.~\ref{MVO Bandgap}. With the increase in pressure, the band gap initially increases up to around 4 GPa, where the system is known to undergo a V coordination change from 5+1 to 6.

Now we shall discuss the optoelectronic behaviour near the isosymmetrical transition. In the case of CaV$_2$O$_6$, upon further compression above 17 GPa, the colour of the sample gradually changes from black to red, accompanied by an increase in the transmission intensity. Above 17 GPa, the optical band gap decreases and stabilizes to a value of about 2.05 eV around 22 GPa, as shown in Fig.~\ref{Bandgap}. 
This clear optical bandgap stabilization matches the exact pressure boundary of the polyhedral distortion anomaly (around 20-23 GPa). This direct alignment proves that the high-pressure state in CaV$_2$O$_6$
hosts a coupled structural-electronic transition, wherein lattice strain boundaries directly manipulate the quantum electronic framework of the system. Further calculations of electronic structure and optical conductivity measurements are needed to establish the microscopic origin of these changes under pressure. Finally, we summarize the results of our study in a tabular form Table~III.

\newpage
\begin{table}[H]
\caption{Summary of high-pressure structural, coordination, photoluminescence (PL)/bandgap, and mechanical/electronic phenomena observed across different pressure regimes.}
\label{tab:high_pressure_data}
\begin{tabular}{p{2cm} p{2cm} p{2.5cm} p{3cm} p{4cm}}
\toprule
\bf{Pressure} & \bf{Symmetry} & \bf{Coordination} & \bf{Optical Properties} & \bf{Structural / Mechanical Behavior} \\
\hline
\addlinespace
$0-1.0$ GPa & Ambient $C2/m$ & Distorted VO$_5$ bipyramids & High PL Intensity  ($\sim$ 705 nm) & Anisotropic structural compression along the c-axis. \\
\addlinespace
$1.0 - 5.2$  GPa & Mixed $C2/m + C2$ & Coexisting ${VO}_5$ \& Distorted ${VO}_6$ & Bandgap widens to $\sim 2.6 eV$; Complete PL Quenching; activation of non-radiative decay pathways & First-order structural reconfiguration. \\
\addlinespace
$20.0 - 31.1 $ GPa & Monoclinic $C2$ (Isosymmetric) & Highly Distorted ${VO}_6$ Octahedra & Color shifts (Black $\rightarrow$ Red); Bandgap plateaus & Elastic anomaly/crossover driven by internal strain and anomalous polyhedral distortion. \\
\bottomrule
\end{tabular}
\label{table-3}
\end{table}

\section{Conclusions}

In conclusion, our comprehensive high-pressure study on CaV$_2$O$_6$ utilizing synchrotron x-ray diffraction, Raman spectroscopy, photoluminescence, and UV-Vis absorption spectroscopy reveals a series of pressure-driven structural, vibrational, and optoelectronic transitions. Upon compression, the ambient centrosymmetric monoclinic $C2/m$ phase converts into a non-centrosymmetric $C2$ structure, characterized by a coordination increase of vanadium ions from VO$_5$ bipyramids to VO$_6$ octahedra. Phase coexistence is observed in the pressure range of 1.0 to 5.2 GPa, while the single $C2$ phase stabilizes fully above 5.2 GPa.   Raman investigations track these microscopic instabilities: negative Grüneisen parameters of low-frequency optical phonons indicate structural softening that facilitates the primary transition into $C2$ symmetry. Concurrently, optical measurements show  complete photoluminescence quenching around 1 GPa triggered by non-radiative pathways from transiently distorted polyhedra, along with a reorganization of band edges that initially widens and subsequently narrows the indirect optical gap. High-pressure optical studies on MgV$_2$O$_6$ confirm that local coordination changes serve as the key structural parameter governing optical gap evolution across divalent metavanadates.   At higher pressures (20 - 23 GPa), an isosymmetric transition occurs without altering crystallographic symmetry, manifesting as an anomaly in internal strain, enhanced polyhedral distortion, a sudden softening of compressibility ($B' < 4$), and a plateauing of the optical band gap. These findings highlight the fundamental interplay between local coordination geometry, mechanical stability, and electronic properties in metavanadates under extreme conditions.

\section{Acknowledgments}
Portions of this research were carried out at the light source PETRA-III of DESY, a member of the Helmholtz Association(HGF). Financial support by the Department of Science and Technology (Government of India) provided within the framework of the India@DESY collaboration is gratefully acknowledged. B.G gratefully acknowledges the CSIR, Government of India, for the financial support and IISER Kolkata for providing experimental facilities to carry out the Ph.D. work.

\newpage

\bibliography{mybib}
\bibliographystyle{unsrt}

\newpage 
\section{Figures}



\end{document}